\documentclass{aa}
\usepackage{natbib}
\usepackage{graphicx}

\def\be{\begin{equation}}
\def\ee{\end{equation}}
\def\lesssim{\raisebox{-0.3ex}{\mbox{$\stackrel{<}{_\sim} \,$}}}
\def\gtrsim{\raisebox{-0.3ex}{\mbox{$\stackrel{>}{_\sim} \,$}}}

\begin{document}

\title{Modelling of surface magnetic field in neutron stars: application to radio pulsars}

 \author{Janusz A. Gil
        \inst{1}
        \and
        George I. Melikidze
        \inst{1,2}
        \and
        Dipanjan Mitra
        \inst{3}}

\institute{Institute of Astronomy, University of Zielona G\'ora,
Lubuska 2, 65-265, Zielona G\'ora, Poland \and Center for Plasma
Astrophysics, Abastumani Astrophysical Observatory, Al.Kazbegi
ave. 2a, Tbilisi 380060, Georgia \and Max-Planck Institute for
Radioastronomy, Auf dem H\"ugel 69, D-53121, Bonn, Germany}


\authorrunning{Gil et al.}

\titlerunning{Modelling of surface magnetic field.}

\date{Received / Accepted }

\abstract{We propose a vacuum gap (VG) model which can be applied
uniformly for normal and high magnetic field pulsars. The model
requires strong and non-dipolar surface magnetic field near the
pulsar polar cap. We assume that the actual surface magnetic field
${\bf B_s}$ in pulsars results from a superposition of global
dipole field ${\bf B_d}$ and crust-anchored small scale magnetic
anomaly ${\bf B_m}$. We provide a numerical formalism for
modelling such structures of surface magnetic field and explore it
within the framework of VG model, which requires strong surface
fields $B_s\gtrsim 10^{13}$~G. Thus, in order to increase the
resultant surface field to values exceeding $10^{13}$~G, in low
magnetic field pulsars with $B_d\ll 10^{13}$~G it is required that
$B_m\gg B_d$, with the same polarities (orientations) of ${\bf
B}_d$ and ${\bf B}_m$. However, if the polarities are opposite,
the resultant surface field can be lower than the dipolar surface
component inferred from the pulsar spin-down. We propose that high
magnetic field pulsars (HBPs) with the inferred global dipole
field $B_d$ exceeding the so called photon splitting threshold
$B_{cr}\sim 4\times 10^{13}$~G, can generate observable radio
emission `against the odds', provided that the surface dipolar
magnetic field $B_d$ is reduced below $B_{cr}$ by the magnetic
anomaly $B_m$ of the right strength and polarity. We find that the
effective reduction is possible if the values of $B_d$ and $B_m$
are of the same order of magnitude, which should be expected in
HBPs with $B_d>B_{cr}$. The proposed VG model of radio emission
from HBPs, in which pair production occurs right above the polar
cap, is an alternative to the recently proposed lengthened space
charge limited flow (SCLF) model, in which pair formation front is
located at relatively high altitudes, where the dipole field is
degraded below $B_{cr}$. Our model allows high $B_d$ radio-loud
pulsars not only just above $B_{cr}$ but even above $2\times
10^{14}$~G, which is the upper limit for HBPs within the
lengthened SCLF model. } \maketitle

 \section{Introduction}

The properties of radio emission of typical pulsars strongly
suggest that the magnetic field is purely dipolar, at least at
altitudes $r$ of several stellar radii $R=10^6$~cm, where the
radio emission is expected to originate \citep[e.g.][and
references therein]{kg97,kg98}. However, this may not be a good
description of the structure of the magnetic field at the stellar
surface. In fact, already Ruderman \& Sutherland (RS75) implicitly
assumed that the radius of curvature of field lines above the
polar cap should be about $10^6$~cm, which is inconsistent with
the global dipolar magnetic field. Several authors argued on
theoretical grounds that the magnetic field could be produced by
currents flowing in thin crustal layers of the neutron star, which
would generate non-dipolar fields at the surface
\citep[e.g.][]{bah83,k91,r91,a93,cr93,gu94,mkb99}. A lot of
observational evidence has also been presented. \citet{ps90} and
\citet{betal92,betal95} reported that interpretation of their
analysis of X-ray pulsars suggested small scale magnetic anomalies
on the polar cap, which would strongly deviate the surface field
from the purely dipolar configuration. It is believed that thermal
X-rays from the polar cap surface are good diagnostic tool to
infer the structure of the surface magnetic field. Several similar
arguments in favour of non-dipolar nature of surface magnetic
field can also be found in \citet{bt97,cgz98,rd99,cz99,td95,td96},
and \citet{metal99}, and \citet{tk01}.

\citet{w64} proposed that the magnetic field is the fossil field
of the progenitor star amplified during the collapse and anchored
in the superfluid core of the neutron star. We will assume in this
paper that the magnetic field was also generated in the outer
crust during or shortly after the neutron star was formed by some
unspecified mechanism \citep[e.g. thermomagnetic
instabilities;][]{bah83}. \citet{uli86} showed that in the crustal
model it is only possible to form small scale surface field
anomalies with a typical size of the order of 100 meters.
\citet{gm01} demonstrated that such `sunspot' like magnetic field
structures on the polar cap surface help to sustain VG-driven
radio emission of pulsars. Here we envisage the scenario where the
magnetic field of neutron star is non-dipolar in nature as a
superposition of the fossil field in the core and the crustal
field structures. The crust gives rise to small scale anomalies
which can be modelled by a number of crust anchored dipoles
oriented in different directions \citep[e.g.][]{bah83,a93}. The
superposition of global dipole and local anomaly is illustrated in
Fig.~1, where for clarity of presentation only one local, crust
associated dipole is marked.

Formation of dense electron-positron pair plasma is essential for
pulsar radiation, especially (but not only) at radio wavelengths.
Purely quantum process for magnetic pair production
$\gamma\rightarrow e^-e^+$ is commonly invoked as a source of this
plasma \citep[e.g.][]{s71,rs75}. However, at superstrong magnetic
fields close to the so-called quantum field $B_{q}=4.4\times
10^{13}$~G, the process of free e$^-$e$^+$ pair production can be
dominated by the phenomenon of photon splitting
\citep{a70,bb70,bh98} and/or bound positronium formation
\citep{um95,um96}. While the latter process can reduce the number
of free pairs at magnetic fields $B\gtrsim 0.1B_{q}$
\citep[e.g.][]{bh01}, the former one can entirely suppress the
magnetic pair production at $B\gtrsim 10^{13}$~G, provided that
photons polarized both parallel and perpendicular to local
magnetic field direction can split \citep[e.g.][]{b01,bh01}. This
assumption will be implicitly kept throughout this paper. Under
these circumstances one can roughly define a photon splitting
critical line $B_{cr}\sim B_q$ and expect that there should be no
radio pulsar above this line on the $B_d-P$ diagram, where
$B_d=6.4\times 10^{19}(P\dot{P})^{1/2}$~G is the dipole surface
magnetic field estimated at the pole from the pulsar period $P$
and its derivative $\dot{P}$ \citep{st83,um96}. This death-line is
more illustrative than quantitative. In fact, a number of specific
model dependent death-lines separating radio-loud from radio-quiet
pulsars are available in the literature
\citep{bh98,bh01,zh00a,zh01}. All these slightly period dependent
death-lines cluster around $B_{q}$ on the $B_d-P$ diagram, and
hence the quantum field is conventionally treated as a threshold
magnetic field above which pulsar radio emission ceases. In this
paper we also use this terminology, bearing in mind that the
photon splitting threshold realistically means a narrow range of
magnetic fields around the critical quantum field $B_{q}\sim
4\times 10^{13}$~G, certainly above $10^{13}$~G \citep[see review
by][]{b01}. For convenience, in all numerical examples presented
in Figs.~2-6 and subsequent discussions we assume the threshold
magnetic field $B_{cr}= B_{q}$.

In order to produce the necessary dense electron-positron plasma,
high voltage accelerating region has to exist near the polar cap
of pulsars. Two models of such acceleration regions were proposed:
stationary space charge limited flow (SCLF) models
\citep{saf78,as79,a81} in which charged particles flow freely from
the polar cap, and highly non-stationary vacuum gap (VG) models
\citep{rs75,cr77,cr80,gm01} in which the free outflow of charged
particles from the polar cap surface is strongly impeded. In the
VG models the charged particles accelerate within a height scale
of about polar cap radius of $\sim 10^4$~cm, due to high potential
drop across the gap, while in the SCLF models particles accelerate
within a height scale of a stellar radius $\sim 10^6$~cm, due to
the potential drop resulting from the curvature of field lines
and/or inertia of outstreaming particles. In both models the free
e$^-$e$^+$ pairs are created if the kinematic threshold
$\varepsilon_\gamma\cdot\sin\theta_{\rm t}=2mc^2$ is reached or
exceeded and the local magnetic field is lower than the photon
splitting threshold $B\sim B_{cr}$, where
$\varepsilon_\gamma=\hbar\omega$ is the photon energy and
$\theta_{\rm t}$ is the propagation angle with respect to the
direction of the local magnetic field.

Recent discovery of high magnetic field pulsars (HBPs) however has
challenged the existing pair creation theories. A few HBPs found
the inferred surface dipolar fields above the photon splitting
level: PSRs J1119$-$6127, J1814$-$1744 and J1726$-$3530 (Table 1).
Moreover, yet another strong field neutron star PSR J1846$-$0258
with $B_d\sim 5\times 10^{13}$~G was discovered \citep{getal00},
which seems to be radio-quiet \citep{ketal96}, although its X-ray
emission is apparently driven by dense $e^-e^+$ pair plasma
\citep[e.g.][]{c01}. However, one should keep in mind that the
actual threshold due to photon splitting and/or bound positronium
formation can be well below the critical field $B_{cr}\sim 4\times
10^{13}$~G, indicating that all high magnetic field radio pulsars
with $B_d>10^{13}$~G pose a challenge. To evade the photon
splitting problem for these pulsars \citet[][ZH00
hereafter]{zh00a} proposed ``a unified picture for HBPs and
magnetars''. They argued that radio-quiet magnetars cannot have
active inner accelerators (thus no $e^-e^+$ pair production),
while the HBPs can, with a difference attributed to the relative
orientations of rotation and magnetic axes (neutron stars can be
either parallel rotator (PRs) with ${\bf\Omega}{\bf B}_d>0$ or
antiparallel rotator (APRs) with ${\bf\Omega}{\bf B}_d<0$, where
${\mathbf\Omega}$ is the pulsar spin axis and $ {\bf B}_d$ is the
magnetic field at the pole). If the photon splitting suppresses
completely the pair production at the polar cap surface, then the
VG inner accelerator cannot form, since the high potential drop
cannot be screened at the top of the acceleration region. Hence,
ZH00 argued that in high magnetic field regime $(B_d>B_{cr})$ the
pair production process is possible only if the SCLF accelerator
forms. In fact, such SCLF accelerators are typically quite long
and their pair formation front (PFF) can occur at high altitudes
$r$, where the dipolar magnetic field $B_d\propto r^{-3}$ has
degraded below the critical value $B_{cr}$. Furthermore, ZH00
demonstrated that such lengthened SCLF accelerator in magnetar
environment can form only for PRs and not for APRs. Consequently
they concluded that the radio-loud HBPs are PRs with developed
lengthened SCLF accelerator, while the radio-quiet magnetars (AXPs
and SGRs) represent APRs. It is worth emphasizing here that ZH00
developed their model under the assumption that the magnetic field
at the surface of HBPs is purely dipolar.

\begin{table}

\caption{Radio-loud HBPs with inferred magnetic field
$B_d=6.4\times 10^{19}(P\dot{P})^{1/2}$~G higher than critical
quantum field $B_{cr}=4.4\times 10^{13}$~G \citep[after Table 1
in][]{zh00b}}
\begin{tabular}{llll}
\hline source & $P$ (s) & $\dot{P}$ (s/s) & $B_d$ (G)\\ \hline
PSR J1814$-$1744 & 3.98 & $7.43\times 10^{-13}$ & $1.1\times 10^{14}$ \\
PSR J1119$-$6127 & 0.41 & $4.02\times 10^{-12}$ & $8.2\times 10^{13}$ \\
PSR J1726$-$3530 & 1.11 & $1.22\times 10^{-12}$ & $7.4\times 10^{13}$ \\
\hline \end{tabular}
\end{table}

In this paper we propose an alternative model for radio-loud HBPs
based on highly non-dipolar surface magnetic field, in which the
photon splitting within the VG inner acceleration region does not
operate even if the dipole magnetic field exceeds the critical
value $B_{cr}$ at the polar cap. Thus our model requires that HBPs
are APRs, which is a consequence of the VG scenario
\citep[e.g.][]{rs75,gm01}. This model is a follow-up work of
\citet{gm01}, who argued that the VG can form if the actual
surface magnetic field is about $10^{13}$~Gauss. In other words,
they assumed that all VG-driven radio pulsars have a very strong,
highly non-dipolar surface magnetic field, with a strength more or
less independent of the value of the global dipole field inferred
from the magnetic breaking law. Thus, if $B_d\ll 10^{13}$~G then
$B_s\gg B_d$ and if $B_d\gtrsim 10^{13}$~G then $B_s\sim B_d$;
however, in any case $B_s<B_{cr}\sim B_q$. We argue that such
strong surface field anomalies can increase low dipolar field
$B_d\ll B_{cr}$ in normal pulsars to values exceeding $10^{13}$~G
(required by conditions for VG formation - see Gil \& Mitra 2001)
if global and local surface fields have the same polarities, or
reduce very high dipolar field $B_d\gtrsim B_{cr}$ in HBPs if both
these components are of comparable values and have opposite
polarities.

\section{Modelling the Surface Magnetic Field}

\begin{figure}
{\includegraphics[width=7cm]{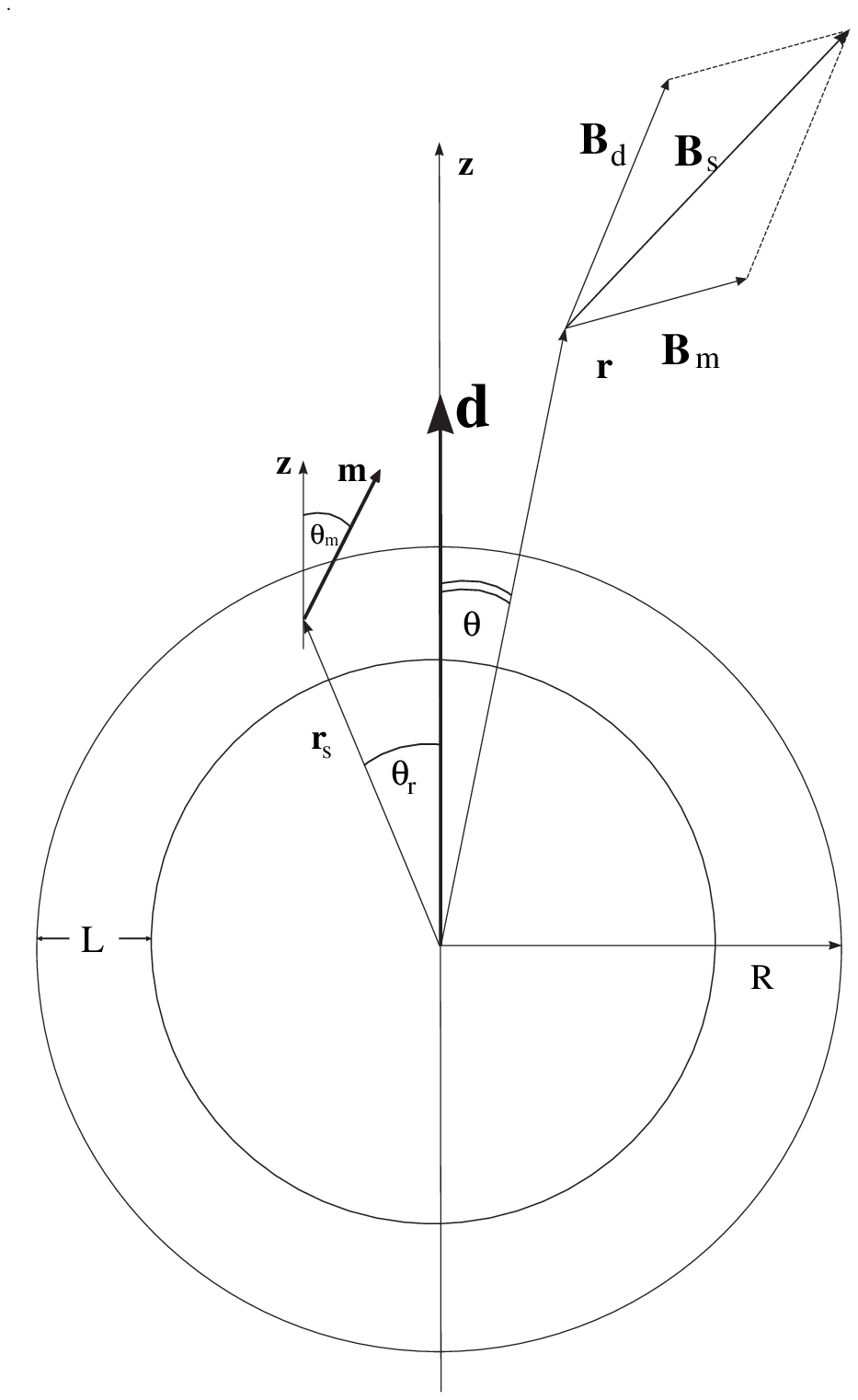}} \caption[]{Superposition of
the star centered global magnetic dipole ${\bf d}$ and crust
anchored local dipole ${\bf m}$ placed at ${\bf r}_s=(r_s\sim
R,0=\theta_r)$ and inclined to the $z$-axis by an angle
$\theta_m$. The actual surface magnetic field at radius vector
${\bf r}=(r,\theta)$ is ${\bf B}_s={\bf B}_d+{\bf B}_m$, where
${\bf B}_d=2{\bf d}/r^3$, ${\bf B}_m=2{\bf m}/|{\bf r}-{\bf
r}_s|$, $r$ is the radius (altitude) and $\theta$ is the polar
angle (magnetic colatitude). $R$ is the radius of the neutron star
and $L$ is the crust thickness. } \label{fig1}
\end{figure}

We model the actual surface magnetic field by  superposition of
the star centered global dipole ${\bf d}$ and a crust anchored
dipole moment ${\bf m}$, whose influence results in small scale
deviations of surface magnetic field from the global dipole.
According to general situation presented in Fig.~\ref{fig1}

\begin{equation}
\mathbf{B}_{s}=\mathbf{B}_{d}+\mathbf{B}_{m},
\end{equation}
where
\begin{equation}
\mathbf{B}_{d}=\left( \frac{2d\cos \theta }{r^{3}},~\frac{d\sin \theta }{%
r^{3}},~0\right),
\end{equation}
$r$ and $\theta $ are star-centered polar co-ordinates, and
\begin{equation}
\mathbf{B}_{m}=\frac{3\left( \mathbf{r}-\mathbf{r}_{s}\right)
\left( \mathbf{m\cdot }\left( \mathbf{r}-\mathbf{r}_{s}\right)
\right) -\mathbf{m} \left|
\mathbf{r}-\mathbf{r}_{s}\right|^{2}}{\left|
\mathbf{r}-\mathbf{r}_{s}\right|^{5}}.
\end{equation}

The global magnetic dipole moment $d=(1/2)B_{d}R^{3}$, where
$B_{d}=6.4\cdot 10^{19}(P\cdot \dot{P})^{1/2}$~G is the dipole
component at the pole derived from the pulsar spin-down rate, and
the crust anchored local dipole moment $m=(1/2)B_{m}\Delta R^{3}$,
where $\Delta R\sim 0.05R$ is the characteristic crust dimension
($R=10^{6}$~cm). We use spherical coordinates with $z$ axis
directed along the global magnetic dipole moment ${\mathbf d}$.
Thus, ${\mathbf{r}}_{s}=(r_{s},\theta _{r},\phi _{r})$ and
${\mathbf{m}}=(m,\theta _{m},\phi _{m})$.

To obtain the equation of the open magnetic field lines we first
define the boundary of the open field lines at an altitude where
the magnetic field should be a pure dipole and we have chosen as a
starting altitude $r=50R$. Then we solve the system of
differential equations
\begin{equation}
\frac{d\theta }{dr}=\frac{B_{\theta }^{d}+B_{\theta }^{m}}{r\left(
B_{r}^{d}+B_{r}^{m}\right) }\equiv \Theta _{1},
\end{equation}
\begin{equation}
\frac{d\phi }{dr}=\frac{B_{\phi }^{m}}{r\left(
B_{r}^{d}+B_{r}^{m}\right) \sin \theta }\equiv \Phi _{1},
\end{equation}
with the initial conditions ${\bf B}_m=0$ defined at $r=50R$
(already at $r=5R$ the ratio $B_m/B_d\sim 10^{-4}$) and trace the
field lines down to the stellar surface ($r=R$). Here
\begin{eqnarray}
B_{r}^{m} &=&-\frac{1}{D^{2.5}}\left(
3Tr_{r}^{s}-3Tr+Dm_{r}\right),
\nonumber \\
B_{\theta }^{m} &=&-\frac{1}{D^{2.5}}\left( 3Tr_{\theta
}^{s}+Dm_{\theta }\right),  \nonumber \\
B_{\phi }^{m} &=&-\frac{1}{D^{2.5}}\left( 3Tr_{\phi }^{s}+Dm_{\phi
}\right)
\end{eqnarray}
and
\begin{eqnarray}
D &=&r_{s}^{2}+r^{2} \nonumber \\
 &-&2r_{s}r\left( \sin \theta_{r}\,
 \sin \theta \,\cos \left( \phi -\phi_{r}\right) +\cos
\theta_{r}\,\cos \theta \right),  \nonumber
\\
T &=&m_{r}r -\left( m_{r}r_{r}^{s}+m_{\theta }r_{\theta
}^{s}+m_{\phi
}r_{\phi }^{s}\right),  \nonumber \\
r_{r}^{s} &=&r_{s}\left( \sin \theta_{r}\,\sin \theta \,\cos
\left( \phi
-\phi_{r}\right) +\cos \theta_{r}\,\cos \theta \right),  \nonumber \\
r_{\theta }^{s} &=&r_{s}\left( \sin \theta_{r}\,\cos \theta \,\cos
\left(
\phi -\phi_{r}\right) -\cos \theta_{r}\,\sin \theta \right),  \nonumber \\
r_{\phi }^{s} &=&-r_{s}\sin \theta_{r}\,\sin \left( \phi
-\phi_{r}\right)
\nonumber \\
m_{r} &=&m\left( \sin \theta _{m}\,\sin \theta \,\cos \left( \phi
-\phi
_{m}\right) +\cos \theta _{m}\,\cos \theta \right),  \nonumber \\
m_{\theta } &=&m\left( \sin \theta _{m}\,\cos \theta \,\cos \left(
\phi
-\phi _{m}\right) -\cos \theta _{m}\,\sin \theta \right),  \nonumber \\
m_{\phi } &=&-m\sin \theta _{m}\,\sin \left( \phi -\phi
_{m}\right).
\end{eqnarray}

The curvature $\rho_c =1/\Re $ of the field lines (where $\Re$ is
the radius of curvature presented for various cases in
Fig.\ref{fig8}) is calculated as
\begin{equation}
\rho_c =\left( \frac{ds}{dr}\right) ^{-3}\left| \left(
\frac{d^{2}\mathbf{r}}{
dr^{2}}\frac{ds}{dr}-\frac{d\mathbf{r}}{dr}\frac{d^{2}s}{dr^{2}}\right)
\right|,
\end{equation}
or
\begin{equation}
\rho_c =\left( S_{1}\right) ^{-3}\left(
J_{1}^{2}+J_{2}^{2}+J_{3}^{2}\right) ^{1/2},
\end{equation}
where
\begin{eqnarray}
\frac{ds}{dr} &=&\sqrt{\left( 1+r^{2}\Theta _{1}^{2}+r^{2}\Phi
_{1}^{2}\sin
^{2}\theta \right) },  \nonumber \\
J_{1} &=& X_{2}S_{1}-X_{1}S_{2},\;\;\; J_{2} =
Y_{2}S_{1}-Y_{1}S_{2}, \nonumber \\
J_{3} &=& Z_{2}S_{1}-Z_{1}S_{2}   \nonumber \\
X_{1} &=& \sin\theta\cos\phi +r\Theta_{1}\cos \theta \cos \phi
-r\Phi_{1}\sin \theta \sin \phi ,  \nonumber \\
Y_{1} &=&\sin \theta \sin \phi +r\Theta_{1}\cos \theta \sin \phi
+r\Phi_{1}\sin \theta \cos \phi ,   \nonumber \\
Z_{1} &=&\cos \theta -r\Theta _{1}\sin \theta ,  \nonumber \\
X_{2} &=& \left( 2\Theta_{1}+r\Theta_{2}\right) \cos \theta \cos
\phi-\left( 2\Phi_{1}+r\Phi _{2}\right) \sin \theta \sin \phi  \nonumber \\
&-&r\left( \Theta_{1}^{2}+\Phi_{1}^{2}\right) \sin \theta \cos
\phi -2r\Theta_{1}\Phi _{1}\cos \theta \sin \phi ,  \nonumber \\
Y_{2} &=& \left( 2\Theta_{1} + r\Theta_{2}\right) \cos \theta \sin
\phi+\left( 2\Phi _{1}+r\Phi _{2}\right) \sin \theta \cos \phi \nonumber \\
&-&r\left(\Theta_{1}^{2}+\Phi_{1}^{2}\right) \sin \theta \sin \phi
+2r\Theta_{1}\Phi _{1}\cos \theta \cos \phi , \nonumber \\
Z_{2} &=& -\Theta_{1}\sin \theta -\Theta_{1}\sin \theta
- r\Theta_{2}\sin \theta -r \Theta_{1}^{2}\cos \theta ,  \nonumber \\
S_{1} &=&\sqrt{ 1+r^{2}\Theta_{1}^{2}+r^{2}\Phi_{1}^{2}\sin^{2}\theta } , \nonumber \\
S_{2} &=&S_{1}^{-1}\left( r\Theta_{1}^{2}+r^{2}\Theta_{1}\Theta
_{2}+r\Phi_{1}^{2}\sin ^{2}\theta \right.  \nonumber \\
&\;&\;\;\;\;\;\;\;+\left. r^{2}\Phi_{1}\Phi_{2}\sin^{2}\theta +
r^{2}\Theta_{1}\Phi_{1}^{2}\sin \theta \cos \theta \right) ,  \nonumber \\
\Theta_{2} &=&\frac{d\Theta_{1}}{dr},\;\;\;\;
\Phi_{2}=\frac{d\Phi_{1}}{dr}.
\end{eqnarray}

For simplicity, in this paper we mostly consider an axially
symmetric case in which both $\mathbf{d} $ and $\mathbf{m}$ are
directed along $z$-axis (parallel or antiparallel), thus
$\theta_{r}=\theta_{m}=\phi_{r}=\phi_{m}=0$. Also, for
convenience, ${\mathbf m}$ is expressed in units of ${\mathbf d}$.
We use normalized units in which $d=P=R=1$ and $r_{s}=0.95$. (see
caption of Fig.~2 for the normalization convention). All
calculations are carried out in three-dimensions, although, for
clarity of graphic presentation, in figures we present only
two-dimensional cuts of the open field line regions.

\subsection{High magnetic field pulsars}

\begin{figure}
{\includegraphics[width=8cm]{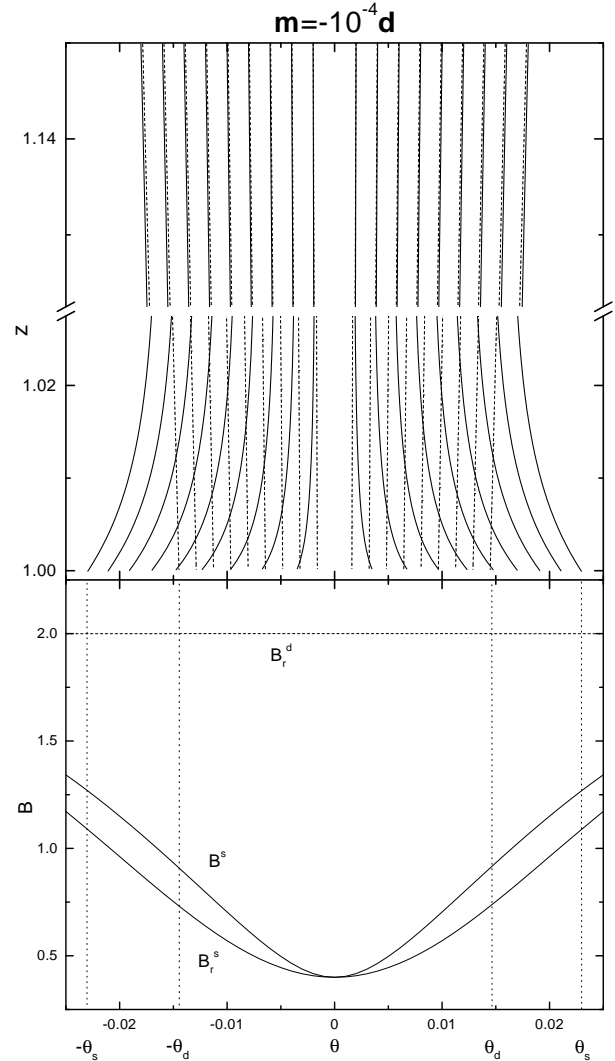}} \caption[]{Structure of
surface magnetic field for a superposition of the global star
centered dipolar moment ${\bf d}$ and crust anchored dipole moment
${\bf m}=-10^{-4}{\bf d}$ (Fig.~\ref{fig1}).  The open dipolar
field lines (solid) and the actual surface open field lines
(dashed) are shown in the upper panel. The horizontal axis is
labelled by an azimuthal angle $\theta$ (magnetic colatitude),
which measures the polar cap radius. For purely dipolar field
lines the polar cap radius $r_d\approx R\cdot\sin\theta_d$, which
for pulsar period $P=1$~s is about 0.014 radians (thus $r_d\approx
1.4\cdot 10^4$~cm). The actual polar cap is broader with the last
open lines emanating at the polar angles $\theta_s\approx 0.023$
(thus the actual polar cap radius $ r_s=2.3\cdot 10^4~{\rm
cm}=1.65r_d$). The actual open surface field lines (solid)
reconnect with dipolar ones (dashed) at distances
$z=(r/R)\cdot\cos\theta\approx 1.2$, where $r$ is the radius and
$\theta<0.025$ radians. In the lower panel the surface values
$(r=R)$ of both dipolar field (dashed horizontal) and the actual
field (solid line) are shown. The radial components
$B_r^d=2d\cos\theta /R^3\approx 2$ and $B_r^s=({\bf B}_d+{\bf
B}_m)\cdot {\bf r}/r$, and total values
$B_s=\sqrt{(B_r^s)^2+(B_\theta^s)^2}$ are presented (where $d=R=1$
is assumed for convenience).}\label{fig2}
\end{figure}

As mentioned above, the formation of VG inner accelerator requires
very high magnetic field $B\gtrsim 10^{13}$~G on the surface of
the polar cap \citep{um95,um96,gm01}. This can be achieved not
only in pulsars with high dipolar field $B_d\gtrsim 10^{13}$~G. In
fact, some of the low field pulsars with $B_d\ll 10^{13}$~G can
have surface field $B_s\gtrsim 10^{13}$~G if $B_m\gg B_d$ (thus
$m\gg 1.25\times 10^{-4}$~d). We discuss such normal, low field
pulsars later in this paper. Presently let us consider the HBP
with a dipolar surface field at the pole $B_d=6.4\times
10^{19}(P\cdot\dot{P})^{1/2}$~G exceeding the photon splitting
limit $B_{cr}\sim B_{q}$. If all photon splitting modes operate,
such pulsar should be radio-quiet. Alternatively these pulsars
could be radio-loud if the effective surface field is reduced
below $B_{cr}$. Such scenario can be achieved if the polarities of
magnetic moments ${\bf d}$ and ${\bf m}$ are opposite, that is
${\bf d}$ and ${\bf m}$ are antiparallel. Fig.~\ref{fig2} presents
a case with ${\bf m}=-10^{-4}{\bf d}$ and $\Delta R/R=0.05$. The
actual surface values of $B^s=\sqrt{(B_r^s)^2+(B_\theta^s)^2}$ as
well as radial components of $B_r^s={\bf B}_s\cdot{\bf R}/R$ and
$B_r^d={\bf B}_d\cdot{\bf R}/R$ are presented in the lower panel
of Fig.~\ref{fig2} (note that all radial components are positive
and that the total $B^s$ is almost equal $B_r^s$ in this case). At
the pole (radius $r=R$ and polar angle $\theta=0$) the ratio
$B_m/B_d=(m/d)\cdot(R/\Delta R)^3=0.8$ and thus
$B_s=B_d-B_m=B_d(1-0.8)=0.2 B_d$. As one can see from this figure,
all surface field lines between $-\theta_s$ and $+\theta_s$ are
open, but the ratio $B_s/B_d$ increases towards the polar cap
edge, reaching the value of about 0.5 in the region between polar
angles $|\theta_d|$ and $|\theta_s|$. The ratio $B_m/B_d$ is also
about 0.5 in this region. Thus, the global dipolar field ($B_d=2$
in our units) is effectively reduced between 2 and 5 times in
different parts of the polar cap (defined as the surface area from
which the open magnetic field lines emanate). This means that the
ratio $B_m/B_d$ ranges from 0.5 to 0.8 across the polar cap. The
actual polar cap is broader than the canonical dipolar polar cap
(two dashed vertical lines correspond to last open dipolar field
lines emanating at the polar angles $\theta_d=\pm 0.014$ radians
for typical period $P=1s$). The ratio of actual to dipolar polar
cap radii is $\theta_s/\theta_d\sim 5/3$ in this case. Thus, using
the argument of magnetic flux conservation of the open field
lines, one can say that an effective surface magnetic field of the
polar cap is about 2.8 times lower than the dipolar surface field
measured from the values of $P$ and $\dot{P}$.  If the estimated
dipole field $B_d^p\approx 10^{14}$~G (like in the case of PSR
J1814$-$1744, Table 1) then the actual surface field at the pole
is only $B_s\sim 2.5\cdot 10^{13}$~G, well below the photon
splitting death line $B_{cr}=4.4\times 10^{13}$~G. Such pulsar can
be radio-loud without invoking the lengthened SCLF accelerator
proposed by ZH00. As shown by \citet{gm01}, in such strong surface
magnetic field the vacuum gap accelerator can form, which implies
low altitude coherent radio emission \citep[][see section 3 in
this paper]{mgp00} at altitudes $r_{em}\sim 50 R$ (for a typical
pulsar with $P=1$~s) in agreement with observational constraints
on radio emission altitudes \citep{c78,c92,kg97,kg98,k01}.

\begin{figure}
{\includegraphics[width=8cm]{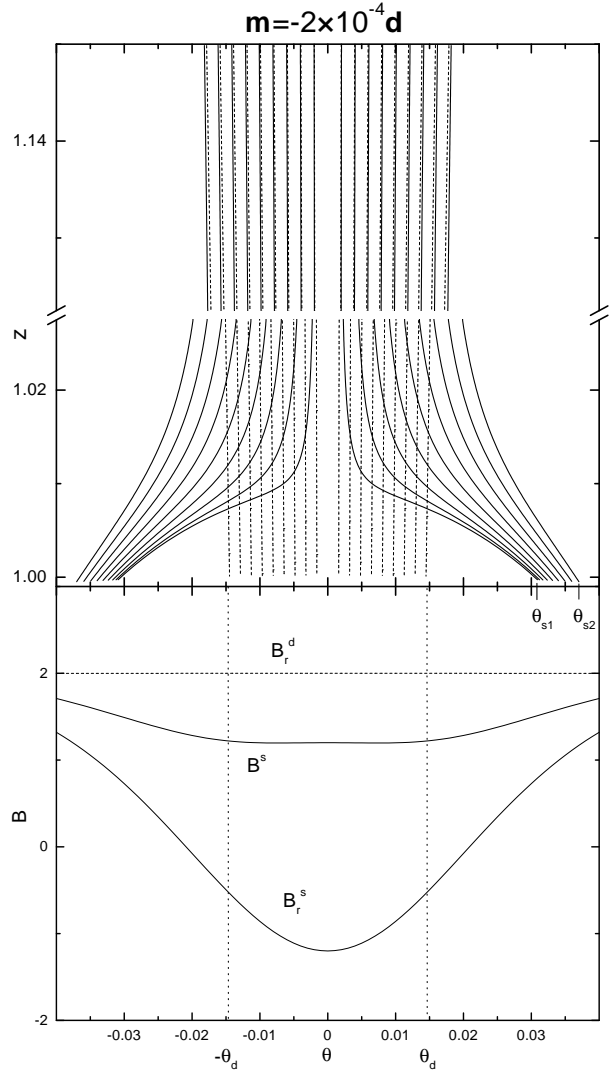}} \caption[]{As in
Fig.~\ref{fig2} but for ${\bf m}=-2\times 10^{-4}{\bf d}$.}
\label{fig3}
\end{figure}

Fig.~\ref{fig3} presents another case of opposite polarities ${\bf
m}=-2\times 10^{-4}{\bf d}$, with a magnitude of ${\bf m}$ two
times stronger than in the previous case (Fig.~\ref{fig2}). Again
for $\Delta R/R\sim 0.05$, $B_m/B_d=(m/d)(R/\Delta R)^3= 1.6$ at
the pole ($r=R$ and $\theta=0$) and $B_s=B_d-B_m=-0.6B_d$. The
negative sign of the ratio $B_s/B_d$ means that the surface
magnetic field ${\bf B}_s$ is directed opposite to ${\bf B}_d$
near the pole, that is the circumpolar field lines with polar
angles $-\theta_{s1}<\theta<\theta_{s1}$ (where $\theta_{s1}\sim
0.031$) are closed. The last open surface field lines (solid)
emanating at polar angles $\theta_s=\pm\theta_{s2}$ (where
$\theta_{s2}\sim 0.037$) reconnect with last open dipolar field
lines (dashed) at altitudes $z \sim 1.2$ (thus about 2 km above
the surface). The actual polar cap, which is the surface through
which the open magnetic field lines emanate, has a shape of a ring
$(0.031\lesssim|\theta_s|\lesssim 0.037)$ located outside the
circle of dipolar polar cap with angular radius $\theta_d=0.014$
(or diameter $r_d\approx\theta_d\cdot R\approx 1.4\cdot 10^4$~cm).
Again, the magnetic flux conservation argument leads to
$B_s/B_d=(\theta_{s_2}^2-\theta_{s_1}^2)/\theta_d^2=(0.037^2-0.031^2)/0.014^2=0.48$,
thus $B_s$ is about $0.5B_d$ within the ring of the open field
lines (thus $|B_m|/B_d\sim 1.6$ in this region. The actual values
of surface magnetic field (radial $B_r^s$ and total
$B^s=\sqrt{(B_r^s)^2+(B_\theta^s)^2}$) are shown as solid lines in
the lower panel of Fig.~\ref{fig3}, in comparison with radial
components of dipolar field $B_d$ (dashed horizontal line). As one
can see, $B_r^s<B_r^d=2m\cos\theta/R^3$ and $B_r^s$ is negative
for $|\theta|<0.022$. If the dipolar surface component of pulsar
magnetic field $B_d\sim 7\times 10^{13}$~G (like in PSR
J1726$-$3530, Table 1), then the actual surface magnetic field
$B_s\sim 4\times 10^{13}$~G, below the photon splitting threshold.

\subsection{Normal pulsars}

\begin{figure}
{\includegraphics[width=8cm]{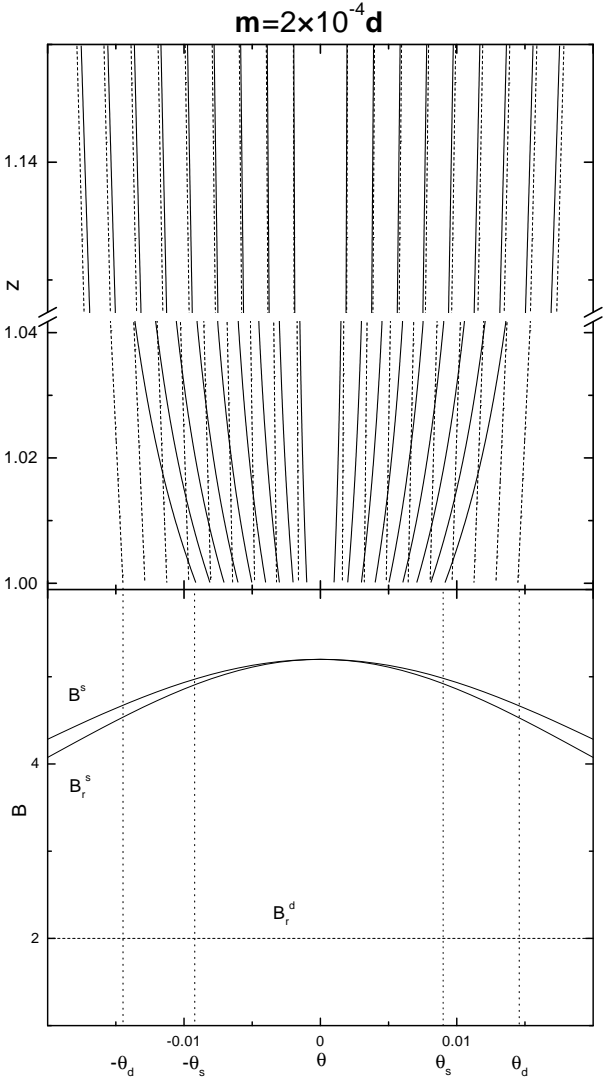}} \caption[]{As in
Fig.~\ref{fig3} but for ${\bf m}=2\times 10^{-4}{\bf
d}$.\label{fig4}}
\end{figure}

Fig. \ref{fig4} presents a case with ${\bf m}=2\times 10^{-4}{\bf
d}$ in which both magnetic moments have the same polarity.
Obviously in such case, the surface magnetic field will be
stronger as compared with pure dipole $(m=0)$. Flux conservation
argument gives surface magnetic field
$B_s/B_d=(\theta_d/\theta_s)^2=(0.014/0.008)^2\sim 3$ (the ration
$B_m/B_d\lesssim 2\times 10^{-4}8000=1.6$). Thus the actual
surface field is about 3 times stronger than the inferred dipolar
field $B_p=6.4\times 10^{19}(P\cdot\dot{P})^{1/2}$~G. It is
interesting to compare this case with the previous one (${\bf
m}=-2\times 10^{-4}{\bf d}$ presented in Fig.~\ref{fig3}), in
which the actual surface field $B_s$ is about 2 times weaker than
the global dipolar surface field at the polar cap. Such cases of
increasing an effective magnetic field can be important in normal
pulsars with low dipolar field $B_d\ll 10^{13}$~G \citep{gm01}.

\begin{figure}
{\includegraphics[width=8cm]{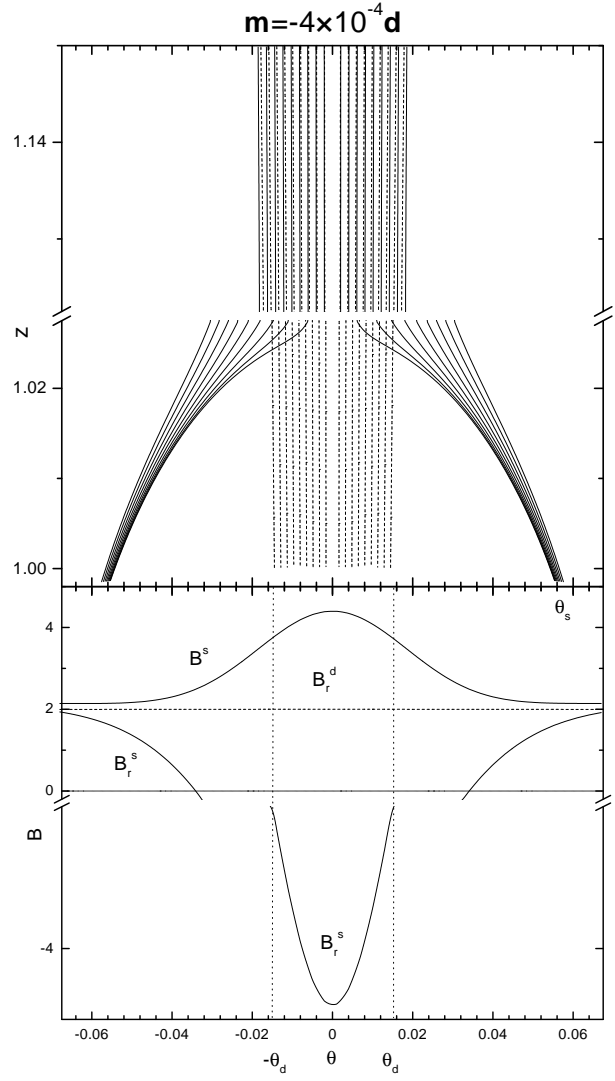}} \caption[]{As in
Fig.~\ref{fig4} but for ${\bf m}=-4\times 10^{-4}{\bf d}$.
\label{fig5}}
\end{figure}

It is then interesting to examine how different polarities of
${\bf d}$ and ${\bf m}$ would influence normal pulsars with
$B_d\ll B_{cr}$. If $m/d\sim(\Delta R/R)^3$ thus $B_m\sim B_d$
then of course $B_s$ can be slightly lower then $B_d$, as in the
case of HBPs (Fig.~\ref{fig2}). In such case, however, the VG
cannot form. In fact, as argued by \citet{gm01}, the formation of
VG requires that $B_s$ is close to $10^{13}$~G or even above, thus
$B_m\gg B_d$ is required in normal pulsars (see also
Gil et al. 2001). Fig.~\ref{fig5}
illustrates a case of high surface magnetic field with $B_m\gg
B_d$, in which VG can apparently form. As one can see from this
figure, the values of $B_s$ at the ring-shaped polar cap are close
to dipolar values $B_s=B_d(r=R,\ |\theta|\sim 0.05)$. One can show
that this is a general situation, that is $B_s\sim B_d$ no matter
how much $B_m$ exceeds $B_d$ at the pole. This follows from the
fact that the angular location $\theta$ of the polar cap ring
increases with the increasing ratio $B_m/B_d\approx (m/d)(\Delta
R/R)^3\simeq 8\times 10^3(m/d)$. For example, in the case
presented in Fig.~\ref{fig5} ${\bf m}=-4\times 10^{-4}{\bf d}$
(thus for $\Delta R=0.05R$ we have $B_m\sim 3B_d$ at the pole) and
the last open field lines emanate at polar angles
$\theta_{s}\approx\pm 0.055$ radians, or at polar cap radii
$R_p\sim 6\times 10^4$~cm (for $P=1$~s). Thus, the narrow polar
cap ring is located far from the local dipole ${\bf m}$, whose
influence is weak at this distance. The circumpolar field lines
between polar angles $-0.053$ to $+0.053$ are closed.

Thus, we conclude that the actual pulsar surface magnetic field
$B_s$ can significantly differ (say by an order of magnitude) from
the inferred dipolar field $B_d$ only in the case when the
polarities of the global ${\bf d}$ and local ${\bf m}$ dipole
(Fig.~\ref{fig1}) are the same, as illustrated in Fig.~\ref{fig4}.
If this is the case, then $B_s$ can largely exceed $B_d$, which
seems to be important from the viewpoint of vacuum gap formation
requiring $B\gtrsim 10^{13}$~G \citep[see][]{gm01}. Therefore, in
normal VG driven radio pulsars the polar cap should be circular,
or at least filled - if the axial symmetry does not hold. The
ring-shaped polar cap can occur only in normal pulsars with
$B_d\lesssim B_{cr}$ and in radio-loud HBPs with $B_d\gtrsim
B_{cr}$.

 \citet{gmm01} explored consequences of
the vacuum gap model interpretation for drifting subpulses observed
in PSR B0943$+$10, in which 20 sparks move circumferentially around the
perimeter of the polar cap, each completing one circulation in 37
pulsar periods \citep{dr99,dr01}. \citet{gmm01} considered both
the curvature radiation (CR) and resonant inverse Compton
radiation (ICS), seed photons as sources of electron-positron
pairs and determined the parameter space for the surface magnetic
field structure in each case. For the CR-VG the surface magnetic
field strength $B_{s}>2\times 10^{13}$~G and the radius of
curvature of surface field lines $0.6\times
10^{5}\mathrm{cm}<{\cal R}<1.2\times
10^{5}\mathrm{cm}$, while for the resonant ICS-VG $B_{s}>2\times 10^{13}$%
~G and $10^{6}\mathrm{cm}<{\cal R}<3\times 10^{6}\mathrm{cm}$ (of
course, in both cased $B_{s}<B_{q}\sim 4.4\times 10^{13}$~G). The
CR-VG with such curved surface magnetic field does not seem likely
(although it cannot be excluded), while the ICS-VG gap supported
by the magnetic field structure determined by the parameter space
determined above guarantees a system of 20 sparks circulating
around the perimeter of the polar cap by means of the
$\mathbf{E}\times \mathbf{B}$ drift in about 37 pulsar periods.

Further \citet{gmm01} modelled the magnetic field structure
determined by the ICS-VG parameter space (specified above), using
the numerical formalism developed in this paper. Since
$B_{d}=6.4\times 10^{12}(P\cdot \dot{P})^{1/2}\mathrm{G}=4\times
10^{12}$~G in this case, then to obtain $\mathbf{B}_{s}\sim (2\div
3)\times 10^{13}$~G one needs $B_{m}\gg B_{d}$ and the same
polarity of both components. Following the symmetry suggested by
the observed patterns of drifting subpulses in PSR B0943$+$10, the
local dipole axis was placed at the polar cap center. A number of
model solutions corresponding to $r_s\sim 0.97$ and $m\sim(1\div
2)\times 10^{-4}$~d and satisfying the ICS-VG parameter space, was
then obtained. As a result of this specific modelling
\citet{gmm01} obtained a number of interesting and important
conclusions: (i) The conditions for the formation of the ICS-VG
are satisfied only at peripheral ring-like region of the polar
cap, which can just accommodate a system of 20 $\mathbf{E}\times
\mathbf{B}$ drifting sparks. (ii) The surface magnetic field lines
within the actual gap are converging, which stabilizes the
$\mathbf{E}\times \mathbf{B}$ drifting sparks by preventing them
from rushing towards the pole (as opposed to the case of diverging
dipolar field \citep[e.g.][]{fr82}. (iii) No model solutions with
$B_s\sim (3\div 4)\times 10^{13}$~G and ${\cal R}\sim(0.6\div
1.2)\times 10^5$~cm, could be obtained which corresponding to the
CR-VG parameter space, which in turn favors the ICS-VG in PSR
B0943$+$10.

\section{Discussion and conclusions}

We argue in this paper that a putative presence of strong
non-dipolar magnetic field on the neutron star surface can help to
understand recently discovered radio pulsars with magnetic field
above the photon splitting threshold, as well as to understand
long standing problems of the vacuum gap formation and drifting
subpulse phenomenon. We model the actual surface magnetic field as
the superposition of the global star-centered dipole and local
crust-anchored dipoles ${\bf B}_s={\bf B}_d+\sum_i{\bf
B}_{mi}\approx {\bf B}_d+{\bf B}_{mo}$, where ${\bf B}_{mo}$ is
the local dipole nearest to the polar cap centre (Fig.~1). Such
model is quite general, as it describes the magnetic field
structure even if the star-centered dipole is negliglible at the
star surface. In such a case the surface dipole field ${\bf B}_d$
(inferred from $P$ and $\dot{P}$ measurements) is a superposition
of all crust-anchored dipoles calculated at far distance and
projected down to the polar cap surface according to the dipolar
law.

We propose a model for radio-loud HBPs with high inferred dipolar
magnetic field $B_d>10^{13}$~G, even exceeding the critical value
$B_{cr}\sim 4\times 10^{13}$~G. Given the difficulty that at
strong magnetic field the magnetic pair creation process is
largely suppressed, the puzzling issue remains how these HBPs
produce their $e^-e^+$ pair plasma necessary for generation of the
observable radio emission. \citet{zh00a} proposed a ``lengthened
version'' of the stationary SCLF model of inner accelerator
\citep[e.g.][]{as79}, in which the pair formation front occurs at
altitudes $r$ high enough above the polar cap that $B_d\sim
B_{cr}(R/r)^3$ degrades below $B_{cr}$, thus evading the photon
splitting threshold. Our VG model is an alternative to the
lengthened SCLF model, with pair creation occurring right at the
polar cap surface, even if magnetic field exceeds $B_{cr}$. We
have assumed that the open surface magnetic field lines result in
an actual pulsar from superposition of the star centered global
dipole moment and a crust anchored local dipole moment. We argued
that if the polarities of these two components are opposite, and
their values are comparable, then the actual value of the surface
magnetic field $B_s$ can be lower than the critical field
$B_{cr}$, even if the global dipole field $B_d$ exceeds the
critical value. Thus, the creation of electron-positron plasma is
possible at least over a part of the polar cap and these high
magnetic field neutron stars can be radio-loud (HBPs). In fact,
one should expect that in HBPs, in which by definition $B_d\gtrsim
B_{cr}\sim B_q=4.4\times 10^{13}$~G, the ratio $B_m/B_d$ should be
of the order of unity, since ${\bf B_s}={\bf B_d}+{\bf B_m}$ and
$10^{13}$~G$\lesssim B_s\lesssim B_{cr}\sim 4\times 10^{13}$~G.

Within our simple model of non-dipolar surface magnetic field
${\bf B}_s$ one should expect that both cases ${\bf m}\cdot{\bf
d}>0$ and ${\bf m}\cdot{\bf d}<0$ will occur with approximately
equal probability. However, from the viewpoint of observable radio
emission only the latter case is interesting in HBPs with
$B_d\gtrsim B_{cr}$. In fact, when ${\bf m}\cdot{\bf d}>0$ then
the surface magnetic field $B_s>B_d\gg B_{cr}$ (Fig.~\ref{fig4})
and the photon splitting level is highly exceeded. For ${\bf
m}\cdot{\bf d}<0$ we have two possibilities: (i) if $m/d\lesssim
(\Delta R/R)^3$ thus $B_m\lesssim B_d$ at the pole $(r=R,
\theta=0)$ then the polar cap (locus of the open field lines) is
circular (Fig.~\ref{fig2}); (ii) if $m/d>(\Delta R/R)^3$ thus
$B_m>B_d$ then part of the circumpolar field lines are closed and
the actual polar cap has the shape of ring (Fig.~\ref{fig3}). In
both above cases (i) and (ii), the actual surface magnetic field
$B_s$ at the polar cap (or at least part of it) can be lower than
$B_{cr}$, even if $B_d$ exceeds $B_{cr}$. The values of $B_m$ and
$B_d\gtrsim B_{cr}$ should be comparable to make reduction of
strong surface field $B_s$ below $B_{cr}$ possible. In our
illustrative examples presented in Figs.~2 and 3 (corresponding to
the same pulsar with $P=1$~s and $B_d=2dR^{-3}=6.4\times
10^{19}(P\cdot\dot{P})^{1/2}$~G) we used ratios $B_m/B_d$ ranging
from 0.5 to 1.6. These values could be slightly different, say by
a factor of few, thus we can say that the ratio $B_m/B_d$ should
be of the order of unity. If $B_m/B_d\gg 1$, then the reduction of
surface dipole field is not effective (see example presented and
discussed in Fig.~5). On the other hand, the case with $B_m\ll
B_d$ is not interesting, as it represents a weak surface magnetic
field anomaly. Thus, among a putative population of neutron stars
with $B_d\gtrsim B_{cr}$, only those with the ratio
$B_m/B_d=(m/d)(R/\Delta R)^3$ of the order of unity, and with
magnetic moment ${\bf m}$ and ${\bf d}$ (Fig.~1) antiparallel at
the polar cap surface, that is ${\bf m}\cdot{\bf d}<0$, can be
detected as HBPs. Other neutron stars from this population of high
magnetic dipole field objects should be radio-quiet. This probably
explains why there are so few HBPs detected.

Within the lengthened SCLF model there is an upper limit around
$B_{d}=2\times10^{14}$~G for radio-loud HBPs (ZH00, Zhang 2001) .
As ZH00 argued, detecting a pulsar above this limit would strongly
imply that only one mode of photon splitting occurs. Without the
alternative model of HBPs proposed in this paper, such detection
would really have great importance for the fundamental physics of
the photon splitting phenomenon. In our VG based model there is no
natural upper limit for the radio-loud HBPs. However, it is known
that due to the magnetic pressure the neutron star surface would
tend to `crack', which should occur at magnetic field strengths
approaching $10^{15}$~G \citep{td95}. It is unclear how the radio
emission would be affected due to such cracking process.

\begin{figure}
{\includegraphics[width=8cm]{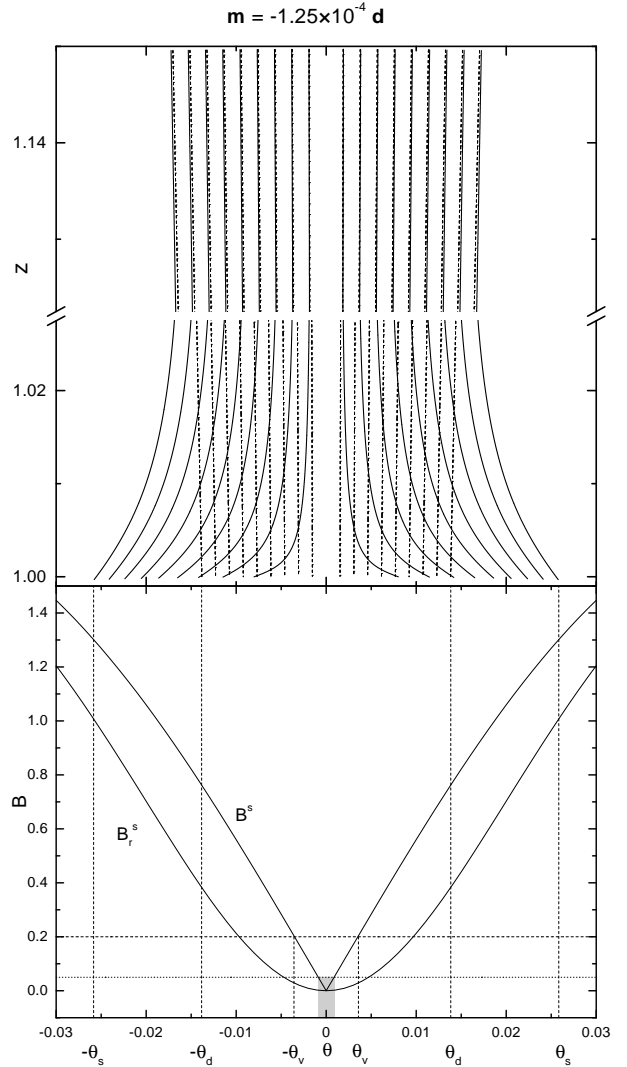}} \caption[]{As in
Fig.~\ref{fig2} but for ${\bf m}=-1.25\times 10^{-4}{\bf d}$. See
also text for explanation. \label{fig6}}
\end{figure}

To illustrate the above argument, let us consider Fig.~\ref{fig6}
which presents yet another case of opposite polarities ${\bf
m}=-1.25\times10^{-4}{\bf d}$. With $\Delta R/R\sim0.05$ this
gives $B_m/B_d=1.0$ and $B_s=B_d-B_m=0$ at the pole ($r=R$,
$\theta=0$). The dashed horizontal line at $B=0.2$ in the lower
panel corresponds to the surface magnetic field $B_s$ which is 10
times weaker than the global dipole component $B_d=2$ (not shown
in the figure). Thus if, for example, $B_d=4\times10^{14}$~G (well
above the lengthened SCLF limit $B_d=2\times10^{14}$~G; such
pulsar was not observed so far), then the actual surface field
$B_s$ is well below $B_{cr}\sim 4\times 10^{13}$~G, at least at
the inner part of the polar cap between $\pm\theta_v=0.0035$~rad.
This ``pair forming effective'' polar cap is about 2.5 times
smaller than the canonical polar cap with radius
$\theta_d=0.014$~rad, and about 7 times smaller than the entire
polar cap with radius $\theta_s=0.027$~rad. Near the last open
field lines at polar angles $0.027\gtrsim|\theta|\gtrsim 0.014$
the actual surface magnetic field $B_s$ is only about 2 times
lower than $B_d$, while in a narrow circumpolar area with
$|\theta|<\theta_v$ the surface field region $B_s$ can be even
more than 10 times weaker than $B_d$. Thus, within our model one
can expect a radio-loud HBP with $B_d$ even exceeding
$4\times10^{14}$~G. However, their radio-beams should be much
narrower than those expected in normal pulsars, at least few to
several times less than $(r_{em}/R)^{1/2}P^{-1/2}$ degrees
\citep[where $r_{em}$ is the radio emission
altitude;][]{kg97,kg98}. This would make such sources difficult to
detect.

\begin{figure}
{\includegraphics[width=8cm]{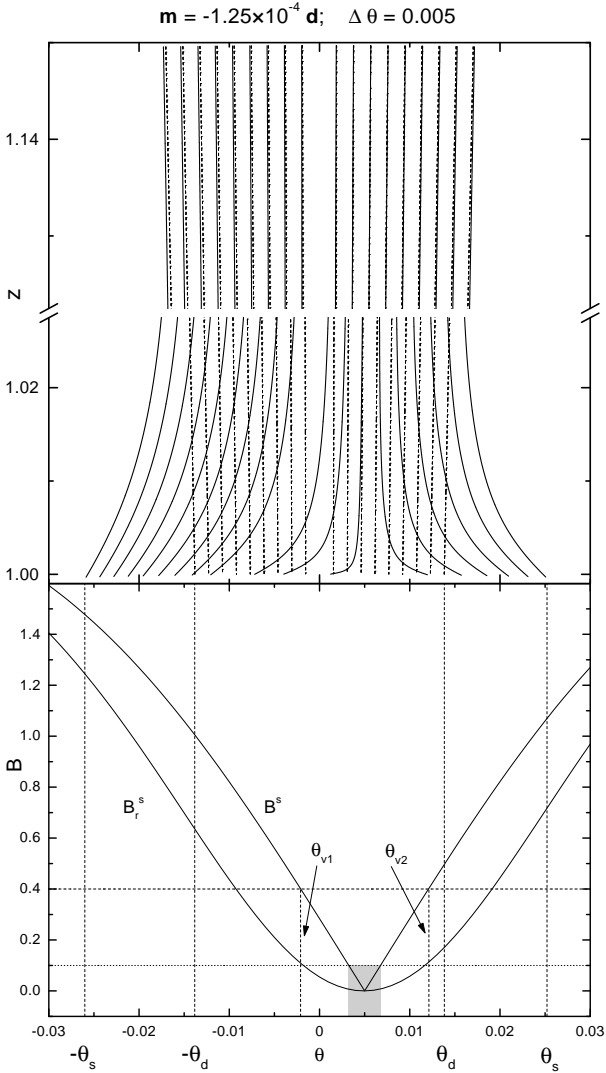}} \caption[]{As in
Fig.~\ref{fig6} but with local dipole shifted off center by
$\Delta\theta=0.005$ radians. See also text for explanation.
\label{fig7}}
\end{figure}

The dotted horizontal line at $B=0.05$ in Fig.~\ref{fig6}
corresponds to $B_s=10^{13}$~G for adopted $B_d=4\times
10^{14}$~G. This value of the surface magnetic field is believed
to be about the lower limit for VG formation
\citep[see][]{gm01,gmm01}. Thus, the shadowed area in
Fig.~\ref{fig6} represent a narrow hollow-cone above which the VG
driven radio emission cannot occur. Similar hollow-cone is marked
in Fig.~\ref{fig7}, which presents the case similar to that
illustrated in Fig.~\ref{fig6}, except the local dipole is shifted
off center by $\Delta\theta=0.005$ radians (corresponding to about
0.2 of the actual polar cap radius). The dashed horizontal line at
$B=0.4$ corresponds to $B_s=4\times 10^{13}$~G and the dotted
horizontal line at $B=0.1$ corresponds to $B_s=10^{13}$~G, both
calculated for adopted $B_d=2\times 10^{14}$~G. The polar angles
$\theta_{v1}$ and $\theta_{v2}$ correspond to $-\theta_v$ and
$+\theta_v$ in Fig.~\ref{fig6}, respectively. The Fig.~\ref{fig7}
demonstrates that conclusions of our paper do not depend on where
the local dipole is placed.

The above arguments strengthen the possibility that some magnetars
can also emit observable radio emission \citep{cetal00,zh00b}. It
is therefore interesting to comment on the apparent proximity of
HBP PSRJ 1814$-$1744 (with $B_p=1.1\times 10^{14}$~G) and AXP 1E
2259$+$586 (with $B_p=1.2\times 10^{14}$~G) on the $P-\dot{P}$
diagram. In both these cases the inferred surface magnetic field
well exceeds the critical value $B_{cr}$. Within our model, the
former object can be radio-loud if the strong local dipole has the
opposite polarity with respect to the global one. The radio
quiescence of the latter object can be naturally explained if the
local dipole is not able to decrease the inferred dipole magnetic
field below the photon splitting death-line. Thus, either the
polarities are the same or they are opposite but the local dipole
is not strong enough to reduce the  dipole surface field below
$B_{cr}$.

\begin{figure}
{\includegraphics[width=8cm]{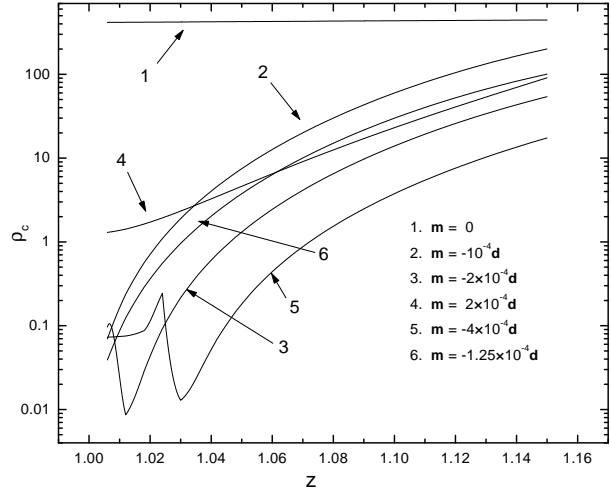}} \caption[]{Radius of
curvature $\rho_c$ (in units of $R=10^6$~cm)  as a function of
normalized altitude $z=(r/R)\cdot\cos\theta$ for actual surface
magnetic field lines corresponding to four cases presented in
Fig.~\ref{fig2}, \ref{fig3}, \ref{fig4}, \ref{fig5} and
\ref{fig6}. For comparison, the radius of curvature of purely
dipolar field lines (in pulsar with P=1 s) is shown (line 1).
\label{fig8}}
\end{figure}

In Fig.~\ref{fig8} we show the radii of curvature of actual
surface field lines compared with those of purely dipolar field
(line 1) as a function of normalized altitude $z=(r/R)\cos\theta$
above the polar cap. Within the polar gap at $ z<1.01$ (within
about 100 meters from the surface) the curvature radii for all
cases presented in Fig.~\ref{fig2}, \ref{fig3}, \ref{fig4},
\ref{fig5} and \ref{fig6} have values of the order of few hundred
meters \citep[see][]{uli86}, suitable for curvature radiation
driven magnetic pair production ($\rho_c=1/\Re<10^6$~cm, where
$\Re$ is the curvature of field lines).

All model calculations performed in this paper correspond to the
axisymmetric case in which one local dipole is placed at the polar
cap center (except the case presented in Fig.~\ref{fig7}). In the
forthcoming paper we will consider a general, non-axisymmetric
case, including more local dipoles, each with different
orientation with respect to the global dipole. Although this
generalization will give more realistic picture of an actual
surface magnetic field, it will not change our conclusions
obtained in this paper.

It should be finally emphasized that although the lengthened SCLF
model for HBPs (ZH00) can solve the problem of pair creation in
pulsars with surface dipole field exceeding the photon splitting
threshold, it does not automatically warrants generation of the
coherent radio emission of such HBPs. The problem is that unlike
in the non-stationary VG model, where the low altitude radio
emission can be generated by means of two-stream instabilities
\citep{am98,mgp00}, the stationary SCLF inner accelerator is
associated with the high altitude relativistic maser radiation
\citep[e.g.][]{kmm91, kmm92, kmm96}. This radiation requires
relatively low Lorentz factors $\gamma_p\sim 5\div 10$ of a dense
secondary plasma \citep[e.g.][]{mu89}. It is not clear if such
plasma can be produced within the lengthened SCLF accelerator with
delayed pair formation taking place in a purely dipolar magnetic
field, either by curvature radiation or by inverse Compton
scattering \citep[e.g.][]{zh00b} processes. Moreover, the
relativistic maser coherent radio emission requires a relatively
weak magnetic field in the generation region. With the surface
dipole field $B_d\sim 10^{14}$~G, such low field may not exist at
reasonable altitudes (about 50\% of the light cylinder radius
$R_L=cP/2\pi$) required by the physics of corresponding
instabilities \citep{kmm91, kmm92, kmm96}. Thus, if one assumes
that the radio-loud HBPs are driven by the SCLF lengthened
accelerator as proposed by ZH00, they might not be able to
generate observable coherent radio emission. This contradiction
seems to be a challenge for the lengthened SCLF scenario for HBPs.
In our VG based model the low altitude $(r\ll R_L)$ radio emission
of HBPs is driven by just the same mechanism as the one most
probably operating in typical radio pulsars \citep[e.g. soliton
curvature radiation proposed recently by][]{mgp00}. In fact, the
HBPs show apparently normal radio emission, with all properties
typical for characteristic pulsar radiation \citep{cetal00}.

\begin{acknowledgements}

This paper is supported in part by the KBN Grant 2~P03D~008~19 of
the Polish State Committee for Scientific Research. We thank E.
Gil and G. Melikidze Jr. for technical help. DM would like to thank
Institute of Astronomy, University of Zielona G\'ora, for support
and hospitality during his visit to the institute, where this and
the accompanying letter paper was started.

\end{acknowledgements}


\begin{thebibliography}{}

\bibitem[Adler et al.(1970)]{a70} Adler, S.L., Bahcall, J.N., Callan, C.G., \& Rosenbluth, M.N. 1970, Phys. Rev. Lett., 25, 1061
\bibitem[Arons(1981)]{a81} Arons, J. 1981, ApJ, 248, 1099
\bibitem[Arons(1993)]{a93} Arons, J. 1993, ApJ, 408, 160
\bibitem[Arons \& Sharleman(1979)]{as79} Arons, J., \& Sharleman, E.T. 1979, ApJ, 231, 854
\bibitem[Asseo \& Melikidze(1998)]{am98} Asseo, E., Melikidze, G. 1998, MNRAS 301, 59
\bibitem[Baring(2001)]{b01} Baring, G.M. 2001, astro-ph/0106161
\bibitem[Baring \& Harding(1998)]{bh98} Baring, G.M., \& Harding, A.K. 1998, ApJ, 507, L55
\bibitem[Baring \& Harding(2001)]{bh01} Baring, G.M., \& Harding, A.K. 2001, ApJ, 547,
929
\bibitem[Becker \& Tr\"{u}mper(1997)]{bt97} Becker, W., \& Tr\"umper, J. 1997, A\&A, 326, 682
\bibitem[Bhattacharya \& Shukre (1985)]{bs85} Bhattacharya, D., \& Shukre, C. S. 1985, JApA, 6, 233
\bibitem[Bialynicka \& Bialynicki et al.(1970)]{bb70} Bialynicka-Birula, Z., \& Bialynicki-Birula, I. 1970, Phys. Rev., 2, 2341
\bibitem[Blandford, Applegate \& Hernquist(1983)]{bah83} Blandford, R.D., Applegate, J.H., \& Hernquist, L. 1983, MNRAS, 204, 1025
\bibitem[Bulik et al.(1992)]{betal92} Bulik, T., Mesarosz, P., Woo, J., Nagase, F., \& Makishima, K. 1992, ApJ, 395, 564
\bibitem[Bulik et al.(1995)]{betal95} Bulik, T., Riffert, H., Mesarosz, P., Makishima, S., \& Mihara, T. 1995, ApJ, 444, 405
\bibitem[Camilo et al.(2000)]{cetal00} Camilo, F., Kaspi, V.M., Lyne, A.G., et al. 2000, ApJ, 541, 367
\bibitem[Chen \& Ruderman(1993)]{cr93} Chen, K., \& Ruderman, M.A. 1993, ApJ, 402, 264
\bibitem[Cheng \& Ruderman(1977)]{cr77} Cheng, K., \& Ruderman, M.A. 1977, ApJ, 214, 598
\bibitem[Cheng \& Ruderman(1980)]{cr80} Cheng, K., \& Ruderman, M.A. 1980, ApJ, 235, 576
\bibitem[Cheng, Gil \& Zhang(1998)]{cgz98} Cheng, K.S., Gil, J., \& Zhang, L. 1998, ApJ, 493, L35
\bibitem[Cheng \& Zhang(1999)]{cz99} Cheng, K.S., \& Zhang, L. 1999, ApJ, 515, 337
\bibitem[Cordes(1978)]{c78} Cordes, J.M. 1978, ApJ, 222, 1006
\bibitem[Cordes(1992)]{c92} Cordes, J.M. 1992, in IAU Coll. 128, ed. T.H. Hankins, J.M. Rankin \& J.A. Gil
Zielona Gora: Pedagogical Univ. Press), 253
\bibitem[Cordes(2001)]{c01} Cordes J.M. 2001, Nature, 409, 296
\bibitem[Deshpande \& Rankin(1999)]{dr99} Deshpande, A.A., \& Rankin, J.M. 1999, ApJ, 524, 1008
\bibitem[Deshpande \& Rankin(2001)]{dr01} Deshpande, A.A., \& Rankin, J.M. 2001, MNRAS, 322, 438
\bibitem[Duncan \& Thompson(1992)]{dt92} Duncan, R., \& Thompson, C. 1992, ApJ, 392, L9
\bibitem[Fillipenko \& Radhakrishnan(1982)]{fr82} Fillipenko,
A.V., \& Radhakrishnan, V. 1982, ApJ, 263, 828
\bibitem[Geppert \& Urpin(1994)]{gu94} Geppert, U., \& Urpin, V.
1994, MNRAS, 271, 490
\bibitem[Gil \& Sendyk(2000)]{gs00} Gil, J., \& Sendyk, M. 2000, ApJ, 541, 351
\bibitem[Gil \& Mitra(2001)]{gm01} Gil, J., \& Mitra, D. 2001, ApJ, 550, 383
\bibitem[Gil et al.(2001)]{gmm01} Gil, J., Melikidze, G.I., \& Mitra, D. 2001, A\&A, submitted
\bibitem[Goldreich \& Julian(1996)]{gj96} Goldreich, P., \& Julian, W.H. 1969, ApJ, 157, 869
\bibitem[Gotthelf et al.(2000)]{getal00} Gotthelf, E.V., Vasisht, G., Boylan-Kolchin, M., \& Torii, A. 2000, ApJ, 542, 37L
\bibitem[Hillebrandt \& M\"{u}ller(1976)]{hm76} Hillebrandt, W., \& M\"{u}ller, E. 1976, ApJ, 207, 589
\bibitem[Kaspi et al.(1996)]{ketal96} Kaspi, N.M., Manchester, R.N., Johnston, S., et al. 1996, Astron. J., 111 (5), 2029
\bibitem[Kazbegi, Machabeli \&  Melikidze(1991)]{kmm91} Kazbegi, A.Z., Machabeli, G.Z., \&  Melikidze, G.I. 1991, MNRAS, 253, 377
\bibitem[Kazbegi, Machabeli \&  Melikidze(1992)]{kmm92} Kazbegi, A.Z., Machabeli, G.Z., \& Melikidze, G.I. 1992, in IAU Coll. 128, ed. T.H. Hankins, J.M. Rankin \& J.A. Gil
Zielona Gora: Pedagogical Univ. Press), 232.
\bibitem[Kazbegi et al.(1996)]{kmm96} Kazbegi, A.Z., Machabeli, G.Z., Melikidze, G.I., et al. 1996, A\&A, 309, 515.
\bibitem[Kijak(2001)]{k01} Kijak, J. 2001, MNRAS, 323, 537
\bibitem[Kijak \& Gil(1997)]{kg97} Kijak, J., \& Gil, J. 1997, MNRAS, 288, 631
\bibitem[Kijak \& Gil(1998)]{kg98} Kijak, J., \& Gil, J. 1998, MNRAS, 299, 855
\bibitem[K\"{o}ssl et al.(1988)]{ketal88} K\"{o}ssl, D., Wolff, R.G., M\"{u}ller, E., \& Hillebrandt, W. 1988, A\&A, 205, 347
\bibitem[Krolik(1991)]{k91} Krolik, J.W. 1991, ApJ, 373, L69
\bibitem[Machabeli \& Usov(1989)]{mu89}Machabeli, G.Z., \& Usov,
V.V. 1989, Sov. Astron. Lett. 15(5), 393
\bibitem[Melikidze, Gil \& Pataraya(2000)]{mgp00} Melikidze, G.I, Gil, J., \& Pataraya, A.D. 2000, ApJ, 544, 1081
\bibitem[Mitra, Konar \& Bhattacharya(1999)]{mkb99} Mitra, D., Konar, S., \& Bhattacharaya, D. 1999, MNRAS, 307, 459
\bibitem[Murakami et al.(1999)]{metal99} Murakami, T., Kubo, S., Shibazaki, N., Takeshima, T., Yoshida, A., \& Kawai,
N. 1999, ApJL, 510, L119
\bibitem[Page \& Sarmiento(1990)]{ps90} Page, D., \& Sarmiento, A. 1996, ApJ, 473, 1067
\bibitem[Pirovaroff, Kaspi \& Camilo(2000)]{pkc00} Pirovaroff, M.J., Kaspi, V.M., \& Camilo, F. 2000, ApJ, 535, 379
\bibitem[Rudak \& Dyks(1999)]{rd99} Rudak, B., \& Dyks J. 1999, MNRAS, 303, 477
\bibitem[Ruderman \& Sutherland(1975)]{rs75} Ruderman, M.A., \& Sutherland, P.G. 1975, ApJ 196, 51
\bibitem[Ruderman(1991)]{r91} Ruderman, M.A. 1991, ApJ, 366, 261
\bibitem[Shapiro \& Teukolsky(1983)]{st83} Shapiro, S.L., \& Teukolsky, S.A. 1983, Black Holes, White Dwarfs and Neutron Stars: The Physics of Compact Objects, New York: Wiley
\bibitem[Sharleman, Arons \& Fawley(1978)]{saf78} Sharleman E.T., Arons J., \& Fawley W.M. 1978, ApJ, 222, 297
\bibitem[Sturrock(1971)]{s71} Sturrock, P.A. 1971, ApJ, 164, 529
\bibitem[Tauris \& Konar(2001)]{tk01} Tauris, T.M., \& Konar, S.
2001, A\&A, 376, 543
\bibitem[Thompson \& Duncan(1995)]{td95} Thompson, C., \& Duncan, R.C. 1995, MNRAS, 275, 255
\bibitem[Thompson \& Duncan(1996)]{td96} Thompson, C., \& Duncan, R.C. 1996, ApJ, 473, 322
\bibitem[Urpin, Levshakov \& Iakovlev (1986)]{uli86} Urpin, V. A., Levshakov, S. A., \& Iakovlev, D. G. 1986, MNRAS, 219, 703
\bibitem[Usov(1987)]{u87} Usov, V.V. 1987, ApJ, 481, L107
\bibitem[Usov \& Melrose(1995)]{um95} Usov, V.V., \& Melrose, D.B. 1995,
Australian J. Phys. 48, 571
\bibitem[Usov \& Melrose(1996)]{um96} Usov, V.V., \& Melrose, D.B. 1996, ApJ, 463, 306
\bibitem[Woltjer (1964)]{w64} Woltjer, L. 1964, ApJ, 140, 1309
\bibitem[Zhang \& Harding(2000a)]{zh00a} Zhang, B., \& Harding, A.K. 2000a, ApJ, 535, L51 (ZH00)
\bibitem[Zhang \& Harding(2000b)]{zh00b} Zhang, B., \& Harding, A.K. 2000b, ApJ, 532, 1150
\bibitem[Zhang \& Harding(2001)]{zh01} Zhang, B., \& Harding, A.K. 2001, Proc. of ``Soft Gamma Repeaters: The Rome 2000 Mini-Workshop'', eds. M.Feroci and S.Mereghetti, Mem.S.A.It.
(ZH01);  astro-ph/0102097
\bibitem[Zhang(2001)]{z01} Zhang, B. 2001, ApJ, 562, in press
\end{thebibliography}
\end{document}